\documentclass[sigconf,authorversion]{acmart}

\usepackage{pbox}
\usepackage{xcolor}
\usepackage[utf8]{inputenc}
\usepackage[show]{chato-notes}
\usepackage{tabu}
\usepackage{caption}
\usepackage{subcaption}
\usepackage[para]{footmisc}
\usepackage[toc,page]{appendix}

\usepackage{listings}
\usepackage{listingsLanguages}
\usepackage{mathtools}
\usepackage{booktabs}
\usepackage{multicol}
\usepackage{multirow}
\usepackage{adjustbox}
\usepackage{hyperref}

\definecolor{maroon}{cmyk}{0, 0.87, 0.68, 0.32}
\definecolor{halfgray}{gray}{0.55}
\definecolor{lightgray}{gray}{0.9}
\definecolor{ipython_frame}{RGB}{207, 207, 207}
\definecolor{ipython_bg}{RGB}{247, 247, 247}
\definecolor{ipython_red}{RGB}{186, 33, 33}
\definecolor{ipython_green}{RGB}{0, 128, 0}
\definecolor{ipython_blue}{RGB}{64, 128, 128}
\definecolor{ipython_purple}{RGB}{170, 34, 255}

\makeatletter
\newcommand\rowtag[2]{#1\def\@currentlabel{#1}\label{#2}}
\makeatother

\usepackage{amsthm}

\usepackage{amsmath}

\newcommand{\crc}[1]{\textcolor{black}{#1}}

\newcommand{\Jpaper}[1]{\textcolor{black}{#1}}
\newcommand{\nicC}[1]{\textcolor{black}{#1}}
\newcommand{\craigC}[1]{\textcolor{black}{#1}}
\newcommand{\cmC}[1]{\textcolor{black}{#1}}

\usepackage{tablefootnote}

\usepackage{array}
\newcolumntype{L}{>{\arraybackslash}m{6cm}}
\newcolumntype{M}{>{\arraybackslash}m{4cm}}
\title[On Approximate Nearest Neighbour Selection for Multi-Stage Dense Retrieval]{On Approximate Nearest Neighbour Selection \\ for Multi-Stage Dense Retrieval}

\author{Craig Macdonald}
\affiliation{%
  \institution{University of Glasgow, UK}
} \email{craig.macdonald@glasgow.ac.uk}

\author{Nicola Tonellotto}
\affiliation{%
  \institution{University of Pisa, Italy}
} \email{Nicola.Tonellotto@unipi.it}

\usepackage{marginnote}
\newcommand{\pageenlarge}[1]{\marginnote{}\enlargethispage{#1\baselineskip}}

\copyrightyear{2021}
\acmYear{2021}
\setcopyright{acmcopyright}\acmConference[CIKM '21]{Proceedings of the 30th ACM International Conference on Information and Knowledge Management}{November 1--5, 2021}{Virtual Event, QLD, Australia}
\acmBooktitle{Proceedings of the 30th ACM International Conference on Information and Knowledge Management (CIKM '21), November 1--5, 2021, Virtual Event, QLD, Australia}
\acmPrice{15.00}
\acmDOI{10.1145/3459637.3482156}
\acmISBN{978-1-4503-8446-9/21/11}


\vspace{-0.34\baselineskip}

\begin{document}

\pagestyle{plain} 
\fancyhead{}
\pagenumbering{gobble}

\begin{abstract}
\looseness -1 Dense retrieval, which describes the use of contextualised language models such as BERT to identify documents from a collection by leveraging approximate nearest neighbour (ANN) techniques, has been increasing in popularity. Two families of approaches have emerged, depending on whether documents and queries are represented by single or multiple embeddings. ColBERT, the exemplar of the latter, uses an ANN index \nicC{and approximate scores} to identify a set of candidate documents for each query embedding, which are then re-ranked using accurate document representations. In this manner, a large number of documents can be retrieved for each query, hindering the efficiency of the \cmC{approach}.  In this work, we investigate the use of ANN scores for ranking the candidate documents, in order to decrease the number of candidate documents being fully scored. \cmC{Experiments conducted on the MSMARCO passage ranking corpus demonstrate that, by cutting of the candidate set by using the approximate scores to only 200 documents, we can still obtain an effective ranking without statistically significant differences in effectiveness, and resulting in a 2$\times$ speedup in efficiency.}

\end{abstract}
\maketitle

\section{Introduction}\label{sec:intro}\pageenlarge{2}

\cmC{Search engine architectures often follow a}
{\em cascading architecture}~\cite{lin2019recant,Matveeva2006HighAR}, in that an initial inexpensive ranking approach (e.g. BM25 applied on a classical inverted index) identifies a candidate set of documents of say $k$ documents, which are then further processed and re-ranked by more expensive ranking approach(es), to identify the final few documents to present to the user. While learning-to-rank with hand-engineered features has typified such a setup, increasingly contextualised language models, such as BERT~\cite{bert}, \cmC{are} used to re-rank the candidate set~\cite{cedr,epic,10.1145/3331184.3331344}. Indeed, BERT has  brought large effectiveness improvements over prior art on information retrieval tasks~\cite{lin2020pretrained}.
\nicC{\cmC{Many} works have proposed different deep network architectures to learn how to represent the semantic content of a sequence of terms, e.g., a document/passage, with a real-value vector, i.e., a \textit{single representation} for the given sequence. This approach collapses all semantic information contained in a given sequence of text into a single vector. A different approach consists \cmC{of} learning a single representation for every token in the sequence, i.e., one per term, generating \textit{multiple representations} for any given text. Several works have shown that multiple representations provide better effectiveness compared to single representations~\cite{khattab2020colbert,polyenc,luan2020}.}

\looseness -1 More recently, effective contextualised language models and approximate nearest neighbour (ANN) techniques~\cite{xiong2020approximate,khattab2020colbert,karpukhin2020dense} have been combined to form {\em dense retrieval}.
\nicC{In dense retrieval, given one or more query embeddings, the most similar documents' embeddings are retrieved using a nearest neighbour search. When single representations for both documents and queries are used, \textit{exact} nearest neighbour (NN) search data structures and algorithms can be exploited~\cite{xiong2020approximate,karpukhin2020dense}. While NN search on single representations has been shown to be efficient, it is less effective than NN search on multiple representations~\cite{lin2020pretrained}. On the other hand, to scale to number of required vectors, multiple representations require \textit{approximate} nearest neighbour (ANN) data structures and algorithms. Hence, Khattab and Zaharia~\cite{khattab2020colbert} propose a \cmC{two-stage} dense retrieval cascading approach. The first stage performs the ANN search to retrieve a set of candidate documents, maximising the recall of the retrieved set. The second stage \cmC{computes accurate scores for} the first-stage candidate documents, to return the final top documents.} \cmC{Overall, compared to re-ranking BM25 results, dense retrieval is appealing as it uses \crc{the} same model end-to-end, therefore improvements to the learned model results in effectiveness improvements in both stages.}



\pageenlarge{0} \looseness -1 \nicC{\cmC{In the ColBERT dense retrieval} approach, the approximate (distance) scores computed in the first stage are ignored in the second stage, as the exact scores for the candidate documents are computed. However, the time spent in the second stage re-ranking is directly proportional to the number of documents retrieved in the first stage.}
In this paper we investigate how rankings can be instantiated upon the approximate first-stage retrieval component, such that a smaller candidate ranking of documents can be passed through to the second stage. In summary, this work contributes an examination on the effectiveness of \crc{ANN} ranking strategies, both alone, and in combination with an effective second stage re-ranker.

\section{Dense Retrieval}\label{sec:densesec}

In this section, we describe a generic architecture for a multi-stage multi-representation dense retrieval system (Section~\ref{ssec:densearch}), as exemplified by the ColBERT approach~\cite{khattab2020colbert}, and discuss how to obtain and use rankings from the first stage of this architecture (Section~\ref{ssec:approx}).


\subsection{Multi-Stage Multi-Representation Dense Retrieval Architecture}\label{ssec:densearch}

\cmC{Interaction-based methods use} multiple embeddings to represent queries and documents, typically one for every query or document term; in models such as BERT these might be WordPieces~\cite{bert}, which represent frequently occurring words or sub-words -- we describe word or-subwords uniformly as {\em tokens}. Therefore queries and documents are represented by a set of fixed-length real-valued vectors, and a similarity score is computed from the query representations and the document representations. Specifically, let a document of \cmC{$|d|$} tokens be represented by a sequence of document embeddings \cmC{$\{\psi_1, \ldots, \psi_{|d|}\}$}, and, similarly, a query of $n$ tokens as query embedding $\phi_i$, i.e., $\{\phi_1, \ldots, \phi_n\}$. Query token embeddings are computed at runtime; queries may also be augmented with additional \emph{masked tokens} to provide which have a query expansion-like role~\cite{khattab2020colbert}.

\begin{figure}[tb]
\includegraphics[width=.99\linewidth]{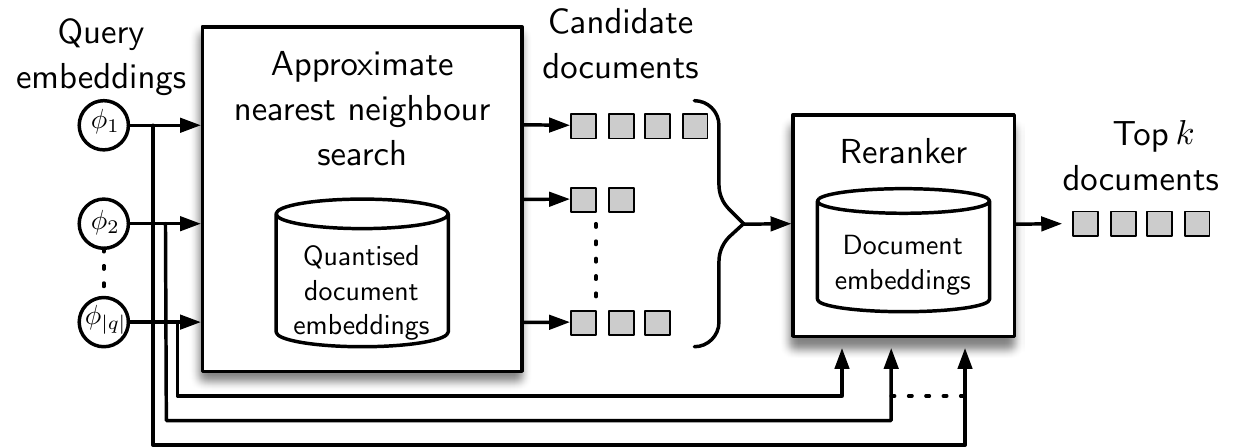}\vspace{-\baselineskip}
\caption{Multi-stage multi-representation dense retrieval.}\label{fig:architecture}\vspace{-\baselineskip}
\end{figure}

\pageenlarge{2} \looseness -1 To support dense retrieval, the general architecture follows a two-stage retrieval cascading architecture. In this setting, a first stage identifies a candidate set of documents using approximate nearest neighbour; in the second stage these documents  are re-ranked by a more expensive exact stage. The overall architecture in shown in Figure~\ref{fig:architecture}. At the time of writing, ColBERT~\cite{khattab2020colbert} is the only effective dense retrieval system exploiting the multiple representations for queries and documents proposed thus far, exhibiting higher effectiveness than dense retrieval based on single query and document representations such as ANCE~\cite{xiong2020approximate} (see~\cite[Table 27 vs.\ Table 28]{lin2020pretrained}).

\subsubsection{Set Retrieval of Query Embedding Approximate Neighbours}

The document embeddings from all documents in the collection are pre-computed and stored into an index data structure for vectors supporting similarity searches. \Jpaper{This can identify the closest vectors to a given input vector when leveraging cosine or dot product vector comparisons.} Indeed, ANN search is applied to compute, for each query embedding $\phi_i$, the set $\Psi(\phi_i, k')$ of the $k'$ document embeddings most similar to $\phi_i$ according to some approximate distance; then, these document embeddings are mapped back to their corresponding documents $D(\phi_i, k')$:
\begin{equation}\label{eq:faiss1}
    D(\phi_i, k') = \{d \in D\;:\; f_D(d) \cap \Psi(\phi_i, k') \neq \emptyset\}
\end{equation}
and finally the union $D(k')$ of these sets is returned:
\begin{equation}\label{eq:faiss2}
    D(k') = \bigcup_{i=1}^{|q|} D(\phi_i, k')
\end{equation}

\looseness -1 \nicC{Indeed, the ANN search can return multiple document embeddings occurring with a single document; the same document can also be retrieved by more than one query embedding, therefore the total returned documents will usually be less than $n \times k'$}. Moreover, some ANN search techniques, e.g.\ those based on product quan\-t\-isat\-ion~\cite{JDH17}, can provide the approximate similarity between query and document embeddings, however, it is not used by ColBERT.


\subsubsection{Exact MaxSim Scoring}

Once the approximate nearest documents $D(k')$ have been identified, they are exploited to compute the final list of top $k$ documents to be returned. To this end, the set of documents $D(k')$ is re-ranked using the query embeddings and the documents' multiple embeddings to produce exact scores, in order to determine the final ranking.

\looseness -1 Given two embeddings, their similarity is computed by the dot product. Hence, for a query $q$ and a document $d$, their final similarity score $s(q,d)$ is obtained by summing up the maximum similarity between the query embeddings and document embeddings:
\begin{equation}\label{eq:maxsim}
    s(q,d) = \sum_{i=1}^{|q|}\max_{j=1,\ldots,|d|} \phi_i^T \psi_j\
\end{equation}

To achieve this, the exact, uncompressed embeddings are stored in a further index structure. Due to the large numbers of documents in the candidate set, the index structure needs to be stored entirely in memory. The MaxSim operator determines the final ranking of documents in response to the query.

\pageenlarge{2} \subsection{Rankings from the Approximate First Stage}\label{ssec:approx}

\nicC{The number of documents in the candidate set produced by ColBERT, denoted $k$, plays an important role.}
Indeed, in the ColBERT instantiation, $k$ is a function of the size of the union of the documents identified by the $k'$ nearest document embeddings to each query embedding. However, allowing each query embedding the same chance to contribute to the candidate set may be sub-optimal. Indeed, consider a query embedding representing a stopword appearing in the query -- retrieving many nearest neighbours to that query embedding is unlikely to retrieve \cmC{as many relevant documents as a more discriminative query embedding}~\crc{\cite{10.1145/3340531.3412079,pruning2021}}\footnote{\crc{Indeed, in \cite{pruning2021}, we show that query embeddings corresponding to tokens with high frequency in the collection can be omitted during nearest neighbours  retrieval.}}.

For this reason, in this paper we investigate different ways to instantiate a ranking of the candidate documents. In doing so, we aim to show that it is possible to (a) have a more precise control \cmC{of} the number of documents requiring further scoring in the second stage ranker, and (b) identify a candidate set that is more effective than a candidate set of a similar size identified using $k'$. In particular, we investigate four methods:

\looseness -1 \textit{Kprime:} This is the default method provided by ColBERT, where the size of the candidate set $k$ is indirectly controlled by $k'$, as described above. The candidate documents are \craigC{unordered}.

\looseness -1 \textit{Count:} As a document is represented by multiple document embeddings, and each document embedding \crc{can be} retrieved by one or more query embeddings, we score documents by the number of times a corresponding document embedding is retrieved in the ANN stage.  \nicC{The candidate documents are ranked by descending count.}

\looseness -1 \textit{Sum Sim:} This method scores \nicC{and ranks} documents by summing the approximate similarities for each query embedding retrieving that document. Approximate similarities are provided \cmC{by the ANN search, but are} not used in the ColBERT implementation. Indeed, they can be inaccurate, due to the ANN quantisation.

\textit{Max Sim:} This method is similar to Sum Sim, but scores \nicC{and ranks} documents by taking for each query embedding the maximum approximate similarity score appearing for any of the document embeddings in the document. In essence, this applies Equation~\eqref{eq:maxsim} at the first stage, but only for \nicC{the} document embeddings that are retrieved for a given query embedding, rather than calculating the full interaction between query and document embeddings.

 \section{Experimental Setup}\label{sec:expsetup}

We experiment to address the following two research questions:

\noindent {\bf RQ1}: How do the approximate ranking methods contribute to an effective candidate set in ColBERT?

\noindent {\bf RQ2}: \crc{How do the use of approximate ranking methods as a first stage} contribute to an effective end-to-end retrieval in ColBERT?


\nicC{In our experiments we use the MSMARCO passage ranking dataset \crc{and the PyTerrier IR experimentation platform~\cite{macdonald2020declarative,pyterrierCIKM}}. We use the ColBERT implementation provided by the authors\footnote{\url{https://github.com/stanford-futuredata/ColBERT/tree/v0.2}}, which we have extended\footnote{\crc{\url{https://github.com/terrierteam/pyterrier_colbert}}}. The general effectiveness of ColBERT is well presented in~\cite{khattab2020colbert}, \cmC{in it which compares} favourably to techniques such as BM25, docT5query~\cite{nogueira2019doc2query} and DeepCT~\cite{dai2020context}.}

\pageenlarge{2} \looseness -1 \cmC{All ColBERT and ANN settings follow~\cite{khattab2020colbert}, namely}: the maximum document length is set to 180 tokens; the maximum query length is set to 32 tokens, including masked tokens; the embeddings size is set to 128, with the resulting document embeddings index being 176 GB; The FAISS ANN index is trained on a randomly selected 5\% sample of the document embeddings, and the resulting IVFPQ index is 16 GB; ANN search is performed on the 10 partitions closest to the input query embedding.

\looseness -1 For evaluating effectiveness, we use the available query sets with relevance assessments: \nicC{6980 queries from the official small version of the MSMARCO Dev set} -- which contain on average 1.1 judgements per query -- as well as the TREC 2019 query set, which contains 43 queries with an average of 215.3 judgements per query. To measure effectiveness, we employ MRR@10 for the MSMARCO Dev set, and the MRR, NDCG@10 and MAP for TREC 2019. We compare our results to the default setting of dense retrieval of ColBERT, using all $32$ query embeddings, and retrieving the default of $k'=1000$ document \cmC{embeddings} for each query embedding. Overall, we \cmC{aim} to maintain high effectiveness \cmC{while reducing the number of documents scored in the second stage compared to the default setting.}




\section{Results}\label{sec:results}

\begin{table*}[tb]
\caption{Effectiveness of first-stage (top half) and end-to-end retrieval (bottom half) when using different approximate ranking methods. * denotes a significant difference in effectiveness compared to the default ColBERT "end-to-end" configuration, according to a paired t-test, $p<0.05$, with Bonferroni multiple testing correction.}\vspace{-0.5\baselineskip}
\centering
\resizebox{138mm}{!}{
\begin{tabular}{ccccccccc}
\toprule
    \multirow{2}{*}{Approach} & \multicolumn{5}{c}{TREC 2019} & \multicolumn{3}{c}{MSMARCO Dev}\\
    \cmidrule(lr){2-6}\cmidrule(lr){7-9}
    & $k$ & MRR & NDCG@10 & MAP & Recall@$k$ & $k$ & MRR@10 & Recall@$k$ \\
    \midrule
    End-to-end    & 7199 & 0.8527 & 0.6934 & 0.3870 & 0.77 & 7151 & 0.344 & 0.97 \\
    \midrule
    \multicolumn{9}{c}{RQ1 - Approximate Only Evaluation} \\
    \midrule
    Count         & 1000 & 0.5346* & 0.3631* & 0.1518* & 0.63*  & 1000 & 0.111* & 0.90*\\
    Approx SumSim & 1000 & 0.4342* & 0.3068* & 0.1226* & 0.62*  & 1000 & 0.084* & 0.90*\\
    Approx MaxSim & 1000 & 0.7425* & 0.5384* & 0.2581* & 0.62*  & 1000 & 0.196* & 0.89*\\
    \midrule
    \multicolumn{9}{c}{RQ2 - Approximate First Stage, Exact Second Stage} \\
    \midrule
    Kprime ($k'=20/100$) & 210 & 0.8488 & 0.6757 & 0.3104* & 0.38* & 901 & 0.341* & 0.85*\\
    Count            & 200 & 0.8365 & 0.6581 & 0.3345* & 0.55* & 1000 & \textbf{0.342}* & \textbf{0.90}*\\
    Approx SumSim    & 200 & 0.8372 & 0.6476 & 0.3199* & 0.52* & 1000 & 0.341* & \textbf{0.90}*\\
    Approx MaxSim    & 200 & \textbf{0.8702} & \textbf{0.6842} & \textbf{0.3487}  & \textbf{0.59}* & 1000 &  \textbf{0.342}* & 0.89*\\
    \bottomrule
\end{tabular}}
\label{tab:rq1}\vspace{-0.5\baselineskip}
\end{table*}

\begin{figure}[tb]
\begin{subfigure}[t]{.49\linewidth}
\includegraphics[width=\linewidth]{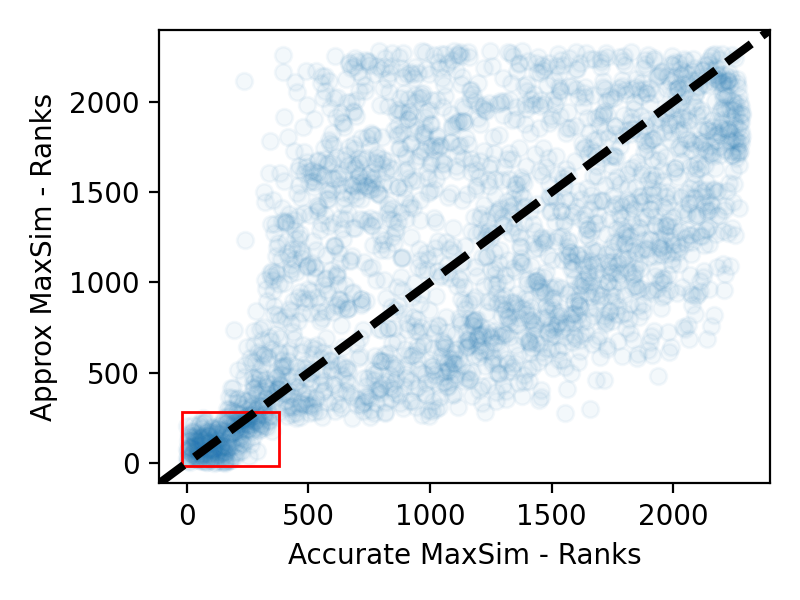}\vspace{-0.5\baselineskip}
\caption{`chemical reactions'}\vspace{-0.5\baselineskip}
\end{subfigure}
\begin{subfigure}[t]{.49\linewidth}
\includegraphics[width=\linewidth]{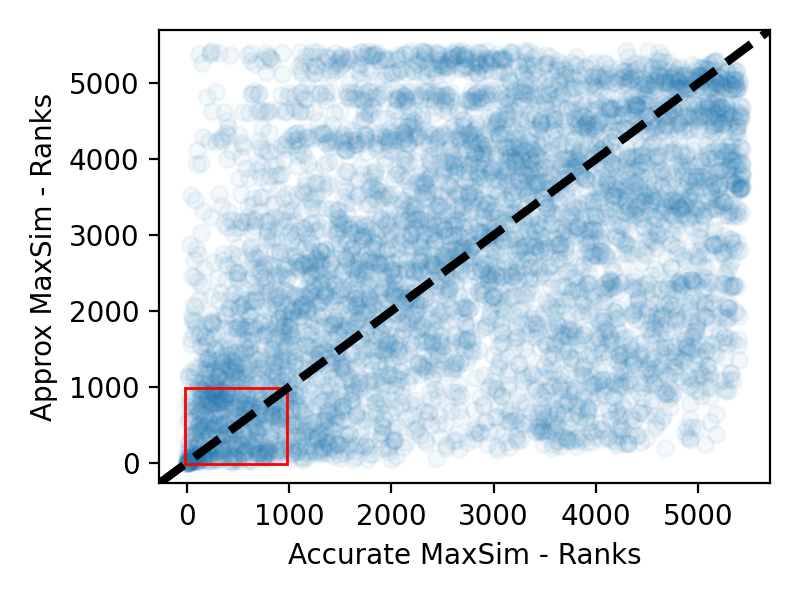}\vspace{-0.5\baselineskip}
\caption{`what is wifi vs bluetooth'}\vspace{-0.5\baselineskip}
\end{subfigure}
\caption{\looseness -1 For two queries, scatterplots comparing accurate ranks of documents vs.\ those from Approx MaxSim. }\label{fig:corr}\vspace{-.5\baselineskip}
\end{figure}

\subsection{RQ1 - Approximate Score Evaluation}
\looseness -1 \cmC{To address} our first research question, we examine the effectiveness of approximate rankings obtained from the ANN index, in comparison to a two stage (``end-to-end'') ranking using the default setting of $k'=1000$. The top half of Table~\ref{tab:rq1} reports the effectiveness on both the TREC 2019 and MSMARCO Dev query sets, in terms of high precision metrics as well as Recall. For approximate ranking methods, we obtain the top-ranked $k=1000$ documents, which is a commonly used setting for a first-stage ranker in LTR settings~\cite{letor,macdonald:2012}. \cmC{Note that Kprime is omitted, as it forms a set, which cannot be evaluated using ranking metrics such as NDCG@10.}

On analysing the top half of  the table, it is easy to observe that the approximate scores obtained from Count, Approx SumSim and Approx MaxSim are insufficiently accurate to obtain high MRR, NDCG or MAP values. Indeed, all metrics are significantly decreased (paired t-test, Bonferroni multiple testing correction), however Approx MaxSim appears to be the most effective. On the other hand, the approximate ranking methods only result in a $\sim$14-15\% reduction in the Recall of relevant documents on the TREC 2019 query set (0.77$\rightarrow$0.63), despite \cmC{reducing the number of retrieved documents by 86\% compared to the default end-to-end configuration.
Similarly, for Dev, Recall is reduced by only 7-8\%, with an analogous reduction in the number of documents retrieved.}
This suggests that applying the approximate ranking methods to identify a smaller candidate set may be a promising direction, which we study further in RQ2.

\pageenlarge{2} To analyse further the accuracy of the approximate methods, Figure~\ref{fig:corr} shows, for two example queries, how the similarity between the rank each document obtained when scored by  the full end-to-end ColBERT pipeline ($x$-axes) compared to that obtained from Approx MaxSim ($y$-axes). Perfect correlation would be exemplified by all points appearing on the \cmC{dashed} $y=x$ \nicC{line}. For the first query, `chemical reactions' (Figure~\ref{fig:corr}(a)), there is only a moderate agreement (\cmC{Spearman's $\rho$=0.57}) about the ranks of the documents obtained when using  exact and approximate scoring methods. 
\nicC{However, a cluster of documents in the top {$\sim$}300 ranks can be observed (see red box) -- both rankings agree about what documents should belong in the top 300, but not \cmC{their relative ordering. We postulate that this cluster may be related to exact matches, which ColBERT scores highly~\cite{10.1007/978-3-030-72240-1_23}.} }
For the longer query `what is wifi vs bluetooth', Figure~\ref{fig:corr}(b), we observe less overall correlation ($\rho=0.40)$, but a little more shading near the origin (see red box), \cmC{which illustrates} more agreement about the highest ranks.

Overall, to answer RQ1, we find that ranking using the approximate scores obtained from the FAISS index result in marked reductions in high precision, but reasonable agreement in which documents should be retrieved (as demonstrated in the Recall reductions, and exemplified by the clusters observed in Figure~\ref{fig:corr}). For this reason, in RQ2, we examine the utility of approximate scoring methods for identifying candidate sets in a two-stage pipeline.

\begin{figure*}[tb]
%
\begin{subfigure}[t]{.295\linewidth}
\includegraphics[width=\linewidth]{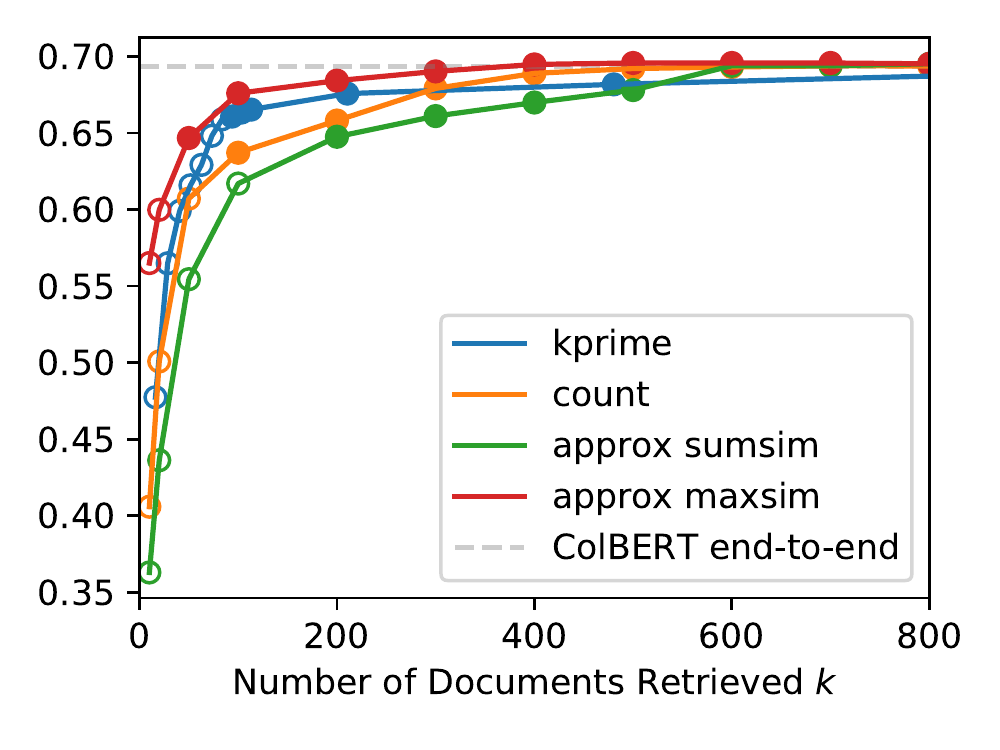}\vspace{-\baselineskip}
\caption{2019, NDCG@10}
\end{subfigure}
\begin{subfigure}[t]{.295\linewidth}
\includegraphics[width=\linewidth]{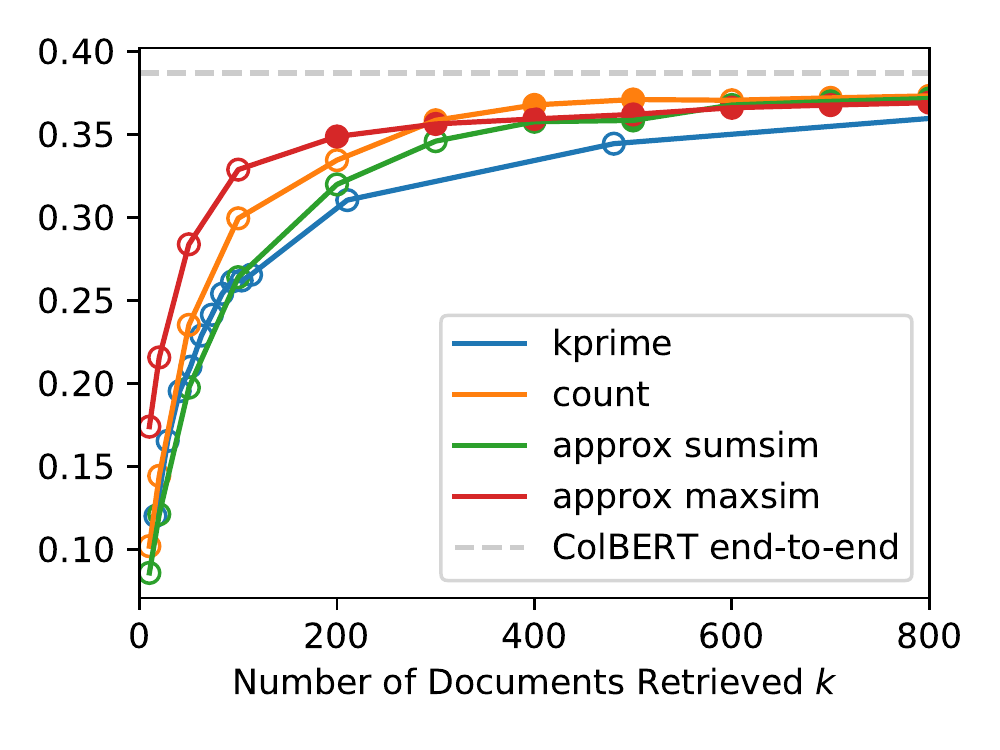}\vspace{-\baselineskip}
\caption{2019, MAP}
\end{subfigure}
\begin{subfigure}[t]{.4\linewidth}
\includegraphics[width=\linewidth]{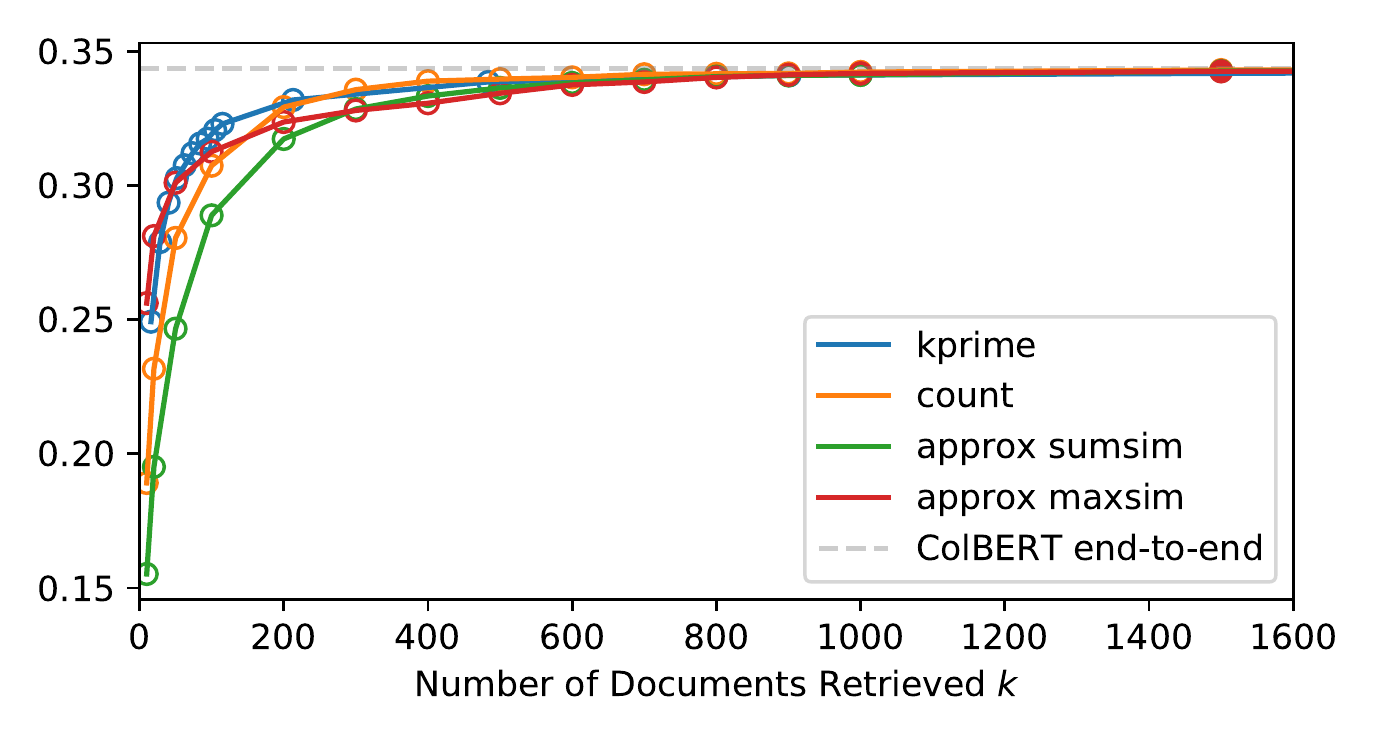}\vspace{-\baselineskip}
\caption{Dev, MRR@10}
\end{subfigure}\vspace{-\baselineskip}
\caption{Effectiveness as $k$ varies. Hollow circles denote statistically significant differences in effectiveness compared to the default ColBERT end-to-end configuration (i.e.\ $k'=1000$), according to a paired t-test, $p<0.05$, Bonferroni correction.}\label{fig:rq2}
\end{figure*}

\subsection{RQ2 - Multi Stage Evaluation}
\looseness -1 Figure~\ref{fig:rq2} plots different retrieval metrics for TREC 2019 and MSMARCO dev query sets, as the number of retrieved documents (i.e. the candidate set size, $k$) varies for different strategies, namely Kprime, Count, Approx SumSim and Approx MaxSim. In particular, for Kprime, we vary $k'$ in [1,10] with step size 1, [10, 50] with step size 10, and [100,1000] with step size 100. For the approximate ranking methods, we control $k$ directly, in $\{10, 20, 50\}$, [100, 1000] with step size 100 and [1000,5000] step size 500. Note that the default ColBERT setting of $k'=1000$ results in, on average, $k\approx 7100$ documents being retrieved for each query and, for legibility, is not shown on the right hand side of the figures. In the figures, hollow circles denote points with statistically indistinguishable effectiveness \cmC{compared to the default ColBERT setting ($k'=1000$)}, according to a paired t-test, $p<0.05$, applying Bonferroni correction.

\looseness -1 On analysing Figure~\ref{fig:rq2}, we observe little degradations in effectiveness until the number of retrieved documents is much smaller than the default (c.f. $k\approx 7100$). Indeed, for NDCG@10 on the TREC 2019 query set, there are no marked degradations in effectiveness until $k \leq 100$; only $k=\{10,20\}$ result in statistically significant decreases in NDCG@10 for Approx MaxSim, and $k=\{10, \ldots, 40\}$ for Approx SumSim. For MAP, upto $k=800$ there is a minor loss in effectiveness compared to the default ColBERT setting (horizontal dashed line), but no significant degradations until $k=300$; Kprime exhibits significant decreases at $k=500$.
On the larger Dev query set, most queries have only a single relevant document, and hence effectiveness decreases more sharply for $k<100$; significant degradations are observed for $k\leq1000$.

\pageenlarge{2} Among the approximate ranking methods, the TREC 2019 query set favours the Approx MaxSim method for NDCG@10, while for Dev, Kprime appears marginally more effective, particularly around $k=100$. Across both query sets, Approx MaxSim appears to be the most effective approximate ranking method, which in essence applies the ColBERT calculation in the earlier first stage.

\cmC{We report $k=200$ for TREC 2019, along with results using $k=1000$ for Dev, in the bottom half of Table~\ref{tab:rq1}. We also report the effectiveness of the Kprime setting that results in $k$ value nearest to 200. Indeed, on TREC 2019, while recall is reduced by 18\% (0.77 $\rightarrow$  0.59), there are no significant differences in MRR, NDCG@10 nor MAP using Approx MaxSim. On Dev, the drop in MRR@10 is statistically significant due to the large number of queries -- we note that at $k=1500$, there is no drop in effectiveness. Overall, for TREC 2019, $k=200$ results in effective retrieval without significant differences in MAP or NDCG@10. Indeed, on our hardware, this resulted in a ${\sim}2\times$ speedup in mean response time (406ms $\rightarrow$ 202ms).}


Hence, in response to RQ2, we find that using the approximate ranking methods allows to control directly the number of retrieved documents, rather than indirectly via $k'$. Applying the MaxSim operator on the approximate scores obtained from the FAISS ANN results in sufficient recall to ensure high effectiveness. For TREC 2019, using $k=200$ with Approx MaxSim resulted in no significant differences in MAP, MRR or NDCG@10, and improved mean response time by a factor of 2.

\vspace{-\baselineskip}\pageenlarge{2}\section{Conclusions}\label{sec:conc}
ColBERT's dense retrieval mechanism can be seen as a first-stage candidate set retrieval, followed by an exact scoring of all the candidates. In this work, we showed than an approximate ranking can be instantiated on the candidate set, and that this allows the size of the candidate set to be markedly reduced (from ${\sim}7100$ documents to $200$), without significantly impacting upon the effectiveness of the final ranking, providing a 2$\times$ speedup in efficiency.

\looseness -1 In this work, we ignored the role of the differing query embeddings used by ColBERT - \cmC{e.g.} retrieving fewer documents for the less important masked query embeddings compared to other query embeddings. We leave this, and \cmC{examining} the impact of varying \cmC{the exact ANN configuration}, to future work.


\section*{Acknowledgements}

Nicola Tonellotto was partially supported by the Italian Ministry of Education and Research (MIUR) in the framework of the CrossLab project (Departments of Excellence). Craig Macdonald acknowledges EPSRC grant EP/R018634/1: Closed-Loop Data Science for Complex, Computationally- \& Data-Intensive Analytics.

\newcounter{BalanceAtReference}
\setcounter{BalanceAtReference}{9}
\newcounter{ReferenceIndexForBalancing}

\makeatletter

\global\@ACM@balancefalse

\def\@balancelastpageonce{%
  \ifnum\value{ReferenceIndexForBalancing}=\value{BalanceAtReference}
    \newpage
  \else
    \relax
  \fi
  \stepcounter{ReferenceIndexForBalancing}
}
\pretocmd{\bibitem}{\@balancelastpageonce}
  {} 
  {\@latex@error{Patching \bibitem failed}{\@ehd}}

\makeatother

\bibliographystyle{ACM-Reference-Format}
\bibliography{bib}






\end{document}